\documentclass{config}

\usepackage[normalem]{ulem}

\usepackage{bbold}
\newtheorem{thm}{Theorem}

\begin{document}

\preprint{APS/123-QED}

\title{Confidence polytopes for quantum process tomography}

\author{E.O. Kiktenko}
\affiliation{Russian Quantum Center, Skolkovo, Moscow 143025, Russia}
\affiliation{Moscow Institute of Physics and Technology, Moscow Region 141700, Russia}
\affiliation{Department of Mathematical Methods for Quantum Technologies, Steklov Mathematical Institute of Russian Academy of Sciences, Moscow 119991, Russia}

\author{D.O. Norkin}
\affiliation{Russian Quantum Center, Skolkovo, Moscow 143025, Russia}
\affiliation{Moscow Institute of Physics and Technology, Moscow Region 141700, Russia}

\author{A.K. Fedorov}
\affiliation{Russian Quantum Center, Skolkovo, Moscow 143025, Russia}
\affiliation{Moscow Institute of Physics and Technology, Moscow Region 141700, Russia}

\date{\today}

\begin{abstract}
	In the present work, we propose a generalization of the confidence polytopes approach for quantum state tomography (QST) to the case of quantum process tomography (QPT).
    	Our approach allows obtaining a confidence region in the polytope form for a Choi matrix of an unknown quantum channel based on the measurement results of the corresponding QPT experiment.
    	The method uses the improved version of the expression for confidence levels for the case of several positive operator-valued measures (POVMs). 
	We then demonstrate how confidence polytopes can be employed for calculating confidence intervals for affine functions of quantum states (Choi matrices), 
	such as fidelities and observables mean values, which are used both in QST and QPT settings.
	As we propose, this problem can be efficiently solved using linear programming tools.
	We also study the performance and scalability of the developed approach on the basis of simulation and experimental data collected using IBM cloud quantum processor. 
\end{abstract}

\maketitle

\section{Introduction}

Quantum tomography is a gold-standard approach for the estimation of the state of a quantum system from measurements~\cite{Lvovsky2009} in experiments with quantum systems of various nature.
Quantum tomography protocols are based on multiple measurements of an unknown quantum state in multiple bases.
The accuracy of quantum state tomography (QST) protocols is limited, in particular, due to statistical errors related to the finite amount of measurement data. 
This poses a problem of estimating the accuracy of QST methods~\cite{Bogdanov2009,Blume-Kohout2012,Renner2012,Flammia2011,Silva2011,Flammia2012,Sugiyama2013,Faist2016,Wang2019}, 
which is of particular importance for characterizing and calibrating quantum devices. 
Existing heuristic tools for solving this problem, such as the numerical bootstrapping method or resampling, are generally biased and lack a well justified error analysis~\cite{Hradil2001,Wineland2009,Tibshirani,Blume-Kohout2012},
which may limit their applicability.

Several rigorously proven methods for QST have been proposed~\cite{Sugiyama2013,Wang2019}.
Specifically, an approach for QST with guaranteed precision has been proposed in Ref.~\cite{Sugiyama2013},
in which the distance between true quantum state and proposed point estimator is provably bounded with high probability.  
The precision-guaranteed approach has been further extended in Ref.~\cite{Wang2019}, where a quantum generalization of Clopper-Pearson confidence intervals~\cite{ClopperPearson1934} has been used. 
Corresponding confidence regions have the shape of a polytope. 
The polytope method is computationally efficient and applicable for experiments~\cite{Wang2019}.

\begin{figure}
	\includegraphics[width=1\linewidth]{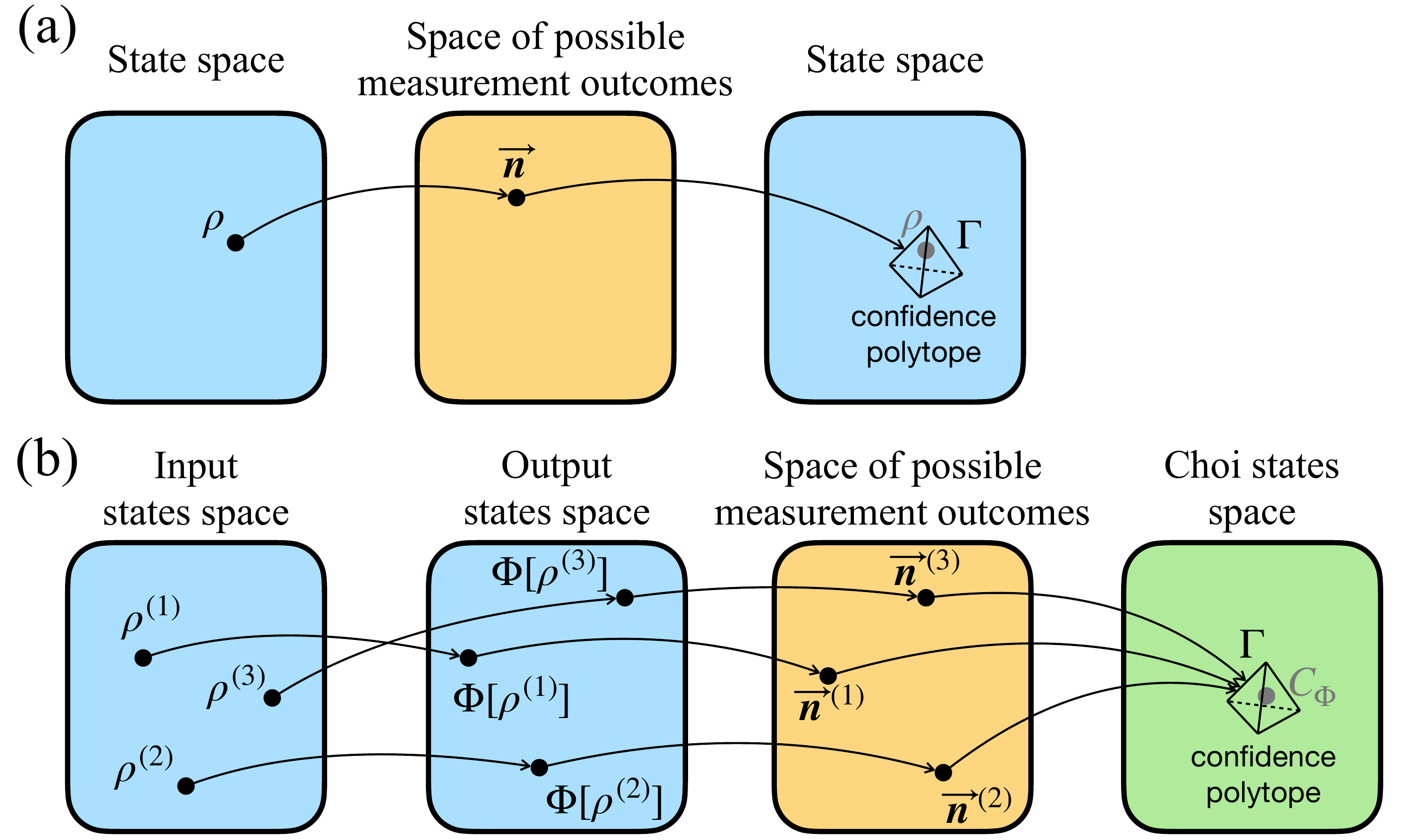}
	\caption{
	Scheme of reconstructing confidence polytopes for (a) quantum state tomography and (b) quantum process tomography.
	A confidence polytope $\Gamma$ defines a set of quantum states (Choi states) such that the reconstructed state $\rho$ (Choi state $C_\Phi$  of reconstructed channel $\Phi$) is among this set at least with probability given by the confidence level.}
	\label{fig:genscheme}
\end{figure} 

The task of obtaining reliable confidence regions is also valuable for quantum process tomography (QPT)~\cite{Poyatos1997,Chuang1997,D'Ariano2001}, which is the natural extension of QST.
QPT can be used for the analysis of the performance of quantum devices and quality of quantum gates in quantum information processing devices~\cite{D'Ariano2002,White2004,Knill2008}.
By exploiting the Choi-Jamiolkowski isomorphism~\cite{Jamiolkowski1972,Choi1975,Jiang2013} also known as the channel-state duality. 
In other words, a quantum channel can be seen as a quantum state so that the QPT task can be reduced to QST of a specific quantum state of higher dimensionality, which is known as the Choi state.
One can then think of using similar methods for estimating the precision of QPT protocols. 
The bootstrapping approach from the QST domain~\cite{Renner2012} has been extended to the QPT case~\cite{Thinh2019}, but without provable accuracy and with high computational costs.  
The generalization of the precision-guaranteed QST scheme~\cite{Sugiyama2013} for quantum processes has also been proposed~\cite{Kiktenko2020}.
Although this approach is computationally efficient, it is limited to the consideration of the Hilbert-Schmidt distance between true Choi state and the corresponding point estimator. 

Here we generalize the polytopes QST approach~\cite{Wang2019} (see Fig.~\ref{fig:genscheme}) that allows one to obtain confidence intervals for various characteristics of quantum processes, such as fidelity and observables mean values. 
We derive confidence regions in the form of a polyhedron (polytope) for a Choi matrix of an unknown quantum processes (channels). 
Specifically, we demonstrate that if the QPT is informationally complete then confidence polyhedrons are bounded, i.e. they are  polytopes.  
In addition, we improve the expression for confidence levels for the case of several positive operator-valued measures (POVMs).
We further derive confidence intervals for affine functions of unknown quantum states and Choi states of unknown processes using linear programming tools and confidence polytopes.
Finally, we show that the accuracy of confidence polytopes for QPT behaves similarly compared to the QST case, 
in particular, the accuracy is almost independent of the Choi matrices space size and number of processed samples in the QPT experiments.
Although we mainly follow the mathematical framework of the estimation theory, the obtained results are directly applicable for characterization and calibration of quantum device in ongoing experiments with physical systems of various nature.
In particular, the considered affine functions include fidelities of reconstructed state with respect to a pure states 
(fidelities of reconstructed process with respect to a unitary operators) and mean values of observables (mean values of observables for outputs state given input state).
We illustrate the practicality of our method by applying it to data from the IBM cloud-accessible quantum processor.

The paper is organized as follows.
In Sec.~\ref{sec:QST}, we revise the confidence polytopes approach for QST and demonstrate how the previously introduced value of the confidence level can be improved in the case of QST with several POVMs.
In Sec.~\ref{sec:QPT}, we introduce the generalization of the confidence polytopes approach for the QPT setting.
In Sec.~\ref{sec:LF}, we demonstrate how to use QST and QPT confidence polytopes for obtaining confidence intervals for affine functions of quantum states (Choi matrices).
In Sec.~\ref{sec:performance}, we study scalability and performance of our method. 
We summarize in Sec.~\ref{sec:conclusion}.

\section{Confidence polytopes for QST} \label{sec:QST}

Here we discuss the method, which was originally proposed in Ref.~\cite{Wang2019}, intended for deriving confidence regions for unknown quantum states based on the measurement protocol and obtained measurement outcomes.
Let us first introduce some basic notations.
We consider a $d$-dimensional ($d$ is finite, $d<\infty$) Hilbert space $\hil$ that is assigned to a quantum system under consideration.
Let $\cal{L}(\hil)$ be a space of linear Hermitian operators over $\hil$, $\mathbb{1}$ be the identity operator in $\cal{L}(\hil)$, ${\cal E}(\hil)=\{E\in {\cal L}(\hil): 0\leq E\leq \mathbb{1}\}$ be a set of measurement effects, 
and $\denset = \left\{\rho\in\mathcal{L}(\hil) : \rho \ge 0,\ \Tr \rho = 1 \right\}$ be a set of density operators.

We then consider a scenario where an unknown state $\rho\in{\cal S}({\cal H})$ is repeatedly prepared and measured with a single POVM $\boldsymbol E=(E_1,\ldots,E_P)$,
which is defined by $P$ effects $E_i\in{\cal E}({\cal H})$ summing up into the identity operator: $\sum_{i=1}^P E_i = \mathbb{1}$.
In what follows it will be important for us to follow the ordering of effects $E_i$, that is why we utilize notation $(\cdot)$ instead of more common $\{\cdot\}$.
Let us denote the obtained measurement outcomes as $\boldsymbol n=(n_1,\ldots,n_P)$, where each integer element $n_i$ indicates the number of appearances of the outcome corresponding to $E_i$.
A method introduced in Ref.~\cite{Wang2019} allows deriving reliable confidence regions for $\rho$ given $\boldsymbol E$ and $\boldsymbol n$.
The confidence region $\Gamma$ is a subset of $\denset$ to which the unknown state $\rho$ belongs with a probability lower-bounded by the value called confidence level ${\rm CL}$:
\begin{equation}
	\Pr{\rho\in\Gamma}>{\rm CL}.
\end{equation}
Here the probability is considered with respect to the measurement outcomes $\boldsymbol n$, which are random variables given the fixed state $\rho$ and POVM $\boldsymbol E$.

\begin{thm}[Confidence region for a single POVM~\cite{Wang2019}] \label{thm:Renner}
	Let an unknown state $\rho\in{\cal S}({\cal H})$ be measured by a POVM $\boldsymbol E=(E_1,\ldots,E_P)$.
	Let
	\begin{multline}\label{eq:effect}
		\Gamma_{\rm eff}(n,N,E,\epsilon)\\:=\left\{\sigma\in\denset: \Tr (\sigma E)\leq \frac{n}{N}+\delta_N(n,\epsilon)\right\},
	\end{multline}
    	where $n$ and $N$ are some integers satisfying $n\leq N$, $E \in {\cal E}({\cal H})$, $\epsilon>0$ is some real positive number, 
    	and $\delta_N(n, \epsilon)$ is a positive root of $D(\frac{n}{N}\|\frac{n}{N} + \delta) = -\frac{1}{N} \log\epsilon$, with $D(x\|y)=x\log(x/y)+(1-x)\log((1-x)/(1-y))$.
	Then for any obtained measurement outcomes $\boldsymbol n=(n_1,\ldots,n_P)$ and a real tuple  $\boldsymbol\epsilon=(\epsilon_1,\ldots,\epsilon_P)$ with $\epsilon_i>0$,
	\begin{equation} \label{eq:POVM}
		\Gamma_{\rm POVM}(\boldsymbol n,\boldsymbol E,\boldsymbol\epsilon):=\bigcap_{i=1}^M \Gamma_{\rm eff}(n_i,\sum_{i=1}^P n_i,E_i,\epsilon_i),
	\end{equation}
	is a confidence region for $\rho$ with confidence level
	\begin{equation}
		{\rm CL}_{\rm POVM}(\boldsymbol\epsilon):=1-\sum_{i=1}^{P}\epsilon_i.
	\end{equation}
\end{thm}
We refer the reader to the original paper~\cite{Wang2019} for the detailed proof.

One can see that the confidence region $\Gamma_{\rm POVM}$ is formed by an intersection of some polyhedron in ${\cal L}({\cal H})$ with a space of density matrices ${\cal S}(\cal H)$.
Let us then discuss a generalization to the cases of several POVMs.

\begin{thm}[Confidence region for a set of POVMs] \label{thm:POVMs}
	Let an unknown state $\rho\in{\cal S}({\cal H})$ be measured by a set of POVMs
	$\vec{\boldsymbol E}=(\boldsymbol E_1,\ldots,\boldsymbol E_L)$, where each $\boldsymbol E_i=(E_{i,1},\ldots E_{i,P_i})$ is a $P_i$-component POVM ($E_{i,j}\in{\cal E}({\cal H})$).
	For any corresponding measurement outcomes $\vec{\boldsymbol n}=({\boldsymbol n}_1,\ldots,{\boldsymbol n}_L)$ with ${\boldsymbol n}_i=(n_{i,1},\ldots,n_{i,P_i})$ 
	and $\vec{\boldsymbol\epsilon}=(\boldsymbol\epsilon_{1},\ldots,\boldsymbol\epsilon_{L})$ 	with $\boldsymbol\epsilon_i=(\epsilon_{i,1},\ldots,\epsilon_{i,P_i})$, $\epsilon_{i,j}>0$,
	\begin{equation} \label{eq:POVMs}
		\Gamma_{\rm POVMs}(\vec{\boldsymbol n}, \vec{\boldsymbol E}, \vec{\boldsymbol\epsilon}):=
		\bigcap_{i=1}^L \Gamma_{\rm POVM}({\boldsymbol n}_i,{\boldsymbol E}_i, \boldsymbol\epsilon_i),
	\end{equation}
	is a confidence region for $\rho$ with the corresponding confidence level
	\begin{equation} \label{eq:CL_POMVs}
		{\rm CL}_{\rm POVMs}(\vec{\boldsymbol\epsilon}):= \prod_{i=1}^L\left(1-\sum_{j=1}^{P_i}\epsilon_{i,j}\right).
	\end{equation}
\end{thm}
See Appendix~\ref{app:thm:POVMs} for the proof.

We note that the introduced confidence level for the several POVMs scenario~\eqref{eq:CL_POMVs} is more tight compared to Ref.~\cite{Wang2019} and given by the following expression:
\begin{equation}
	{\rm CL}'_{\rm POVMs}(\vec{\boldsymbol\epsilon}):=
	1-\sum_{i=1}^L\sum_{j=1}^{P_i}\epsilon_{i,j}.
\end{equation}
This improvement comes from utilizing the independence of measurement outcomes for different POVMs.

Let us summarize the physical essence of Theorems~\ref{thm:Renner} and~\ref{thm:POVMs}. 
The core idea is that within a QST protocol it is possible to bound an area in the state space such that the true unknown state is inside this area with at least certain predetermined probability. 
The area has the shape of a polyhedron intersected with the set of physically possible states.
The polyhedron is determined by the employed measurements (given by POVMs) and the obtained outcomes statistics: 
Each effect of each POVM determines an orientation of a facet plane of the polyhedron, while the number of corresponding measurement outcomes determines the parallel transfer of this plane.
An increase in the number of measurements brings the planes closer to each other and makes the polyhedron less bulky, thus decreasing the uncertainty about the true state.
The resulting confidence level for the case of several POVMs can be obtained as a product of confidence levels related to each particular POVM.

To conclude this section, we consider an embedding of ${\cal S}({\cal H})$ into $\mathbb{R}^{d^2-1}$ (it will be used below for constructing confidence intervals for affine functions of unknown quantum states in Sec.~\ref{sec:LF}).
For this purpose, let us introduce an orthogonal set $\{\sigma_i\}_{i=1}^{d^2-1}$ of traceless Hermitian operators $\sigma_i\in{\cal L}({\cal H})$ satisfying condition $\Tr (\sigma_i \sigma_j)=d\delta_{i,j}$, where $\delta_{i,j}$ stands for Kronecker symbol.
Then for any $\rho \in {\cal S}({\cal H})$ and $E\in{\cal E}({\cal H})$, we can assign vectors ${\boldsymbol r} (\rho)\in\mathbb{R}^{d^2-1}$ and  $\boldsymbol\eta(E)\in\mathbb{R}^{d^2-1}$ with components given by
\begin{equation}
	r_i(\rho) := \Tr (\sigma_i \rho) \quad \mbox{ and } \quad
	\eta_i(E) := \frac{1}{d}\Tr (\sigma_i E)
\end{equation}
respectively.
We also set $\eta_0(E):=d^{-1}\Tr E$, and so for any $\rho\in{\cal S}({\cal H})$ and $E\in{\cal L}(\cal H)$ we have
\begin{equation} \label{eq:probofmeasinvec}
	\Tr (\rho E)={\boldsymbol r}(\rho) \cdot \boldsymbol\eta(E)+\eta_0(E),    
\end{equation}
where $\boldsymbol{x}\cdot \boldsymbol{y}$ stands for the standard dot-product of $(d^2-1)$-dimensional vectors.

Therefore, for the subset $\Gamma_{\rm eff}(n,N,E,\epsilon)\subset {\cal S}({\cal H})$, defined by Eq.~\eqref{eq:effect}, we introduce an analog in $\mathbb{R}^{d^2-1}$ as follows:
\begin{multline}
	\widetilde\Gamma_{\rm eff}(n,N,E,\epsilon):=\\\left\{{\boldsymbol r}\in\mathbb{R}^{d^2-1}: {\boldsymbol r}
	\cdot\boldsymbol{\eta}(E)\leq \frac{n}{N}+\delta_N(n,\epsilon)-\eta_0(E)\right\}.
\end{multline}
Similar expressions can be then introduced for $\Gamma_{\rm POVM}$ and $\Gamma_{\rm POVMs}$ in the following way
\begin{eqnarray}
	&\widetilde\Gamma_{\rm POVM}({\boldsymbol n},{\boldsymbol E},\boldsymbol\epsilon)&:=\bigcap_{i=1}^P \widetilde\Gamma_{\rm eff}(n_i,\sum_{i=1}^Pn_i,E_i,\epsilon_i),\\ &\widetilde\Gamma_{\rm POVMs}(\vec{\boldsymbol n}, \vec{\boldsymbol E},
	\vec{\boldsymbol\epsilon})&:=
        \bigcap_{i=1}^L \widetilde\Gamma_{\rm POVM}({\boldsymbol n}_i,{\boldsymbol E}_i, \boldsymbol\epsilon_i).
\end{eqnarray}
We see that $\widetilde\Gamma_{\rm POVMs}$ is a fair polyhedron in $\mathbb{R}^{d^2-1}$, 
which can also possess points $\boldsymbol{r}(\sigma)$ with unit-trace but not positive semidefinte $\sigma$ (note that $\widetilde\Gamma_{\rm POVM}$ 
can be considered as a special case of $\widetilde\Gamma_{\rm POVMs}$ with $\vec{{\boldsymbol E}}=({\boldsymbol E})$).
Thus, one can think about $\widetilde\Gamma_{\rm POVMs}$ as a result of removing semi-positivity restriction from $\Gamma_{\rm POVMs}$:
\begin{equation} \label{eq:state_extention}
	\rho \in
	\Gamma_{\rm POVMs}(\vec{\boldsymbol n}, \vec{\boldsymbol E},
	\vec{\boldsymbol\epsilon})
	\Rightarrow
	\boldsymbol{r}(\rho)\in
	\widetilde\Gamma_{\rm POVMs}(\vec{\boldsymbol n}, \vec{\boldsymbol E},
	\vec{\boldsymbol\epsilon}),
\end{equation}
the opposite, however, in general is not true.
To conclude the section, we formulate the necessary and sufficient condition for $\widetilde\Gamma_{\rm POVMs}$ to be bounded.
\begin{thm} \label{thm:POVMs-polytope}
	$\widetilde\Gamma_{\rm POVMs}(\vec{\boldsymbol n}, \vec{\boldsymbol E},
	\vec{\boldsymbol\epsilon})$ is a polytope (bounded polyhedron) iff $\vec{\boldsymbol E}$ is informationally complete.
\end{thm}
See Appendix~\ref{sec:thm:POVMs-polytope} for the proof. 
The essence of Theorem~\ref{thm:POVMs-polytope} is that in the case, where employed in QST measurements `cover' the whole state space, 
the resulting confidence polyhedron is actually a polytope, and so is bounded in all directions regardless of the restriction posed by the boundedness of the physical states space.

\section{Confidence polytopes in QPT}\label{sec:QPT}

Here we introduce a generalization of the confidence polyhedrons (polytopes) framework to the case of quantum channels.
For this purpose consider a completely-positive trace-preserving map (CPTP) 
also known as a channel $\Phi\!: {\cal L}({\cal H_{\rm in}})\rightarrow {\cal L}({\cal H_{\rm out}})$, where ${\cal H_{\rm in}}$ and ${\cal H_{\rm out}}$ are $d_{\rm in}$- and $d_{\rm out}$-dimensional Hilbert spaces ($d_{\rm in}, d_{\rm out}<\infty$), respectively.
A common example of a quantum channel is the map connecting the initial state of an open quantum system with its final state after certain period of time, 
given that during this time period the system interacts with an environment prepared in a certain fixed initial state (the initial system-plus-environment state is assumed to be factorized).
In what follows, $\mathbb{1}_{{\rm in}({\rm out})}$ is the identity operator in ${\cal H}_{{\rm in}({\rm out})}$, and $\Tr_{{\rm in}({\rm out})}$ denotes partial trace over ${\cal H}_{{\rm in}({\rm out})}$.

To define the map $\Phi$ it is convenient to consider a corresponding Choi state $C_{\Phi}\in {\cal C}({\cal H}_{\rm in},{\cal H}_{\rm out}):=\{C\in {\cal L}({\cal H}_{\rm in}\otimes{\cal H}_{\rm out}): C\geq0, \Tr_{\rm out} C = \mathbb{1}_{\rm in} \}$, given by
\begin{equation}
C_{\Phi} = \Sum_{i,j=0}^{d_{\rm in}-1} \ket{i}\bra{j}\otimes\Phi\left(\ket{i}\bra{j}\right),
\end{equation}
where $\{\ket{n}\}_{n=0}^{d_{\rm in-1}}$ is the standard computational basis in ${\cal H}_{\rm in}$.
Physically, the Choi state can be interpreted as a (renormalized) `response' of the channel to a given proper maximally entangled state.
Recall that given $C_{\Phi}$ one can compute an output of $\Phi$ for any input $\rho\in{\cal L}(\cal H_{\rm in})$ as follows:
\begin{equation}
    \Phi[\rho] = \Tr_{\rm in}( \rho^\top \otimes \mathbb{1}_{\rm out} C_{\Phi}),
\end{equation}
where $\top$ stands for standard transposition.

Consider a QPT protocol, where a set of input states $\boldsymbol\rho_{\rm in}=(\rho_{\rm in}^{(1)},\ldots,\rho_{\rm in}^{(M)})$ with $\rho_{\rm in}^{(i)}\in{\cal S}({\cal H}_{\rm in})$ is repeatedly prepared, put through $\Phi$, and then measured by a set of POVMs $\vec{\boldsymbol E}^{(i)}=({\boldsymbol E}^{(i)}_1,\ldots, {\boldsymbol E}^{(i)}_{L^{(i)}})$, where each ${\boldsymbol E}_j$ is some $P_j^{(i)}$-component POVM.
The measurement results in this case are given by a dataset $\boldsymbol{\mathfrak{n}} := ({\vec{\boldsymbol n}^{(i)}})_{i=1}^M$, where each $\vec{\boldsymbol n}^{(i)}$ has the structure of QST protocol considered in the previous section.
We then obtain a confidence region in the form
\begin{equation}
	\Pr{C_\Phi \in \Gamma} > {\rm CL},
\end{equation}
based on $\boldsymbol\rho_{\rm in}$, $\boldsymbol{\mathfrak{E}} := ({\vec{\boldsymbol E}}^{(i)})_{i=1}^M$, and $\boldsymbol{\mathfrak{n}}$.

Our main result regarding confidence polytopes for quantum processes (channels) is stated as follows.
\begin{thm}[Confidence region for a CPTP] \label{thm:CPTP}
	Consider a QPT protocol run for an unknown quantum channel $\Phi$ with $C_{\Phi}\in{\cal C}({\cal H}_{d_{\rm in}},{\cal H}_{d_{\rm out}})$, which is specified by a set of input states $\boldsymbol\rho_{\rm in}$
	and measurements of corresponding output states
	$\boldsymbol{\mathfrak{E}}$.
	Let for arbitrary $P$-component POVM ${\boldsymbol E}=(E_i)_{i=1}^{P}$, $E_i\in {\cal E}({\cal H}_{\rm out})$, tuple of non-negative integers ${\boldsymbol n}=(n_i)_{i=1}^P$, state $\rho_{\rm in}\in{\cal S}({\cal H}_{\rm in})$, and tuple of positive real values $\boldsymbol\epsilon=(\epsilon_i)_{i=1}^P$
    \begin{multline} \label{eq:Gamma_ij}
        \Gamma_\text{\rm in-out}({\boldsymbol n}, \rho_{\rm in}, {\boldsymbol E}, \boldsymbol\epsilon) :=
        \left\{
            C\in{\cal C}({\cal H}_{d_{\rm in}},{\cal H}_{d_{\rm out}}):
        \right.\\
        \left.
            \Tr_{\rm in}(\rho_{\rm in}^{\top}\otimes\mathbb{1}_{\rm out} C)\in \Gamma_{\rm POVM}\left({\boldsymbol n}, {\boldsymbol E},\boldsymbol{\epsilon}\right)
        \right\}.
    \end{multline}
    Then for any obtained measurement dataset $\boldsymbol{\mathfrak{n}}$ and a set of real positive values $\boldsymbol\varepsilon= (\vec{\boldsymbol{\epsilon}}^{(i)})_{i=1}^M$,
    $\vec{\boldsymbol\epsilon}^{(i)}=({\boldsymbol\epsilon}^{(i)}_j)_{j=1}^{L^{(i)}}$,
    ${\boldsymbol\epsilon}^{(i)}_j=(\epsilon_{j,k}^{(i)})_{k=1}^{P^{(i)}_j}$, $\epsilon_{j,k}^{(i)}>0$,
    \begin{multline} \label{eq:proc}
       \Gamma_{\rm CPTP}\left(\boldsymbol{\mathfrak{n}},
       \boldsymbol\rho_{\rm in},
       \boldsymbol{\mathfrak{E}},
       \boldsymbol\varepsilon\right)
       :=\\
    \bigcap_{i=1}^{M} \bigcap_{j=1}^{L^{(i)}}
       \Gamma_\text{\rm in-out}\left({\boldsymbol n}^{(i)}_j,
       \rho^{(i)}_{\rm in},
       {\boldsymbol E}^{(i)}_j,
       \boldsymbol\epsilon^{(i)}_{j}\right),
    \end{multline}
    is a confidence region for $C_{\Phi}$ with confidence level
    \begin{equation} \label{eq:CL_proc}
         {\rm CL}_{\rm CPTP}
         \left(\boldsymbol\varepsilon\right)
         :=\prod_{i=1}^{M}\prod_{j=1}^{L^{(i)}}\left(1-\sum_{k=1}^{P^{(i)}_{j,k}}\epsilon^{(i)}_{j,k}\right).
    \end{equation}
\end{thm}

See Appendix~\ref{app:thm:CPTP} for the proof.

One can see that confidence region given by Eq.~\eqref{eq:Gamma_ij} and Eq.~\eqref{eq:proc} is quite straightforward generalization of the one for quantum states, which are given by Eqs.~\eqref{eq:effect}, \eqref{eq:POVM}, and \eqref{eq:POVMs}.
We note that the structure of the confidence level~\eqref{eq:CL_proc} includes products with respect to different input states and POVMs.
In this way, the physical essence of Theorem~\ref{thm:CPTP} is that within a QPT experiment it is possible to bound an area in the space of all possible quantum channels, 
i.e. space of Choi states, such that the reconstructed channel is inside this area with at least certain predetermined probability. 
The obtained confidence region has the form of a polyhedron intersected with the set of physically admissible quantum channels.
The facet planes orientations of the polyhedron are determined by combinations of input states and POVMs effects of output measurements, and the positions of facet planes are determined by the number of corresponding measurement outcomes.
As in the case of QST, the volume of the polyhedron is determined by the amount of accumulated statistics: the more measurements are used, the less uncertainty about the true channel is.

In order to derive confidence intervals for affine function of Choi matrices, 
we consider an embedding of ${\cal C}({\cal H}_{\rm in},{\cal H}_{\rm out})$ into $\mathbb{R}^{d_{\rm in}^2}\times \mathbb{R}^{d_{\rm out}^2-1}$.
For this purpose, we employ two orthogonal sets: $\{\sigma^{\rm in}_i\}_{i=0}^{d_{\rm in}^2-1}$ and
$\{\sigma^{\rm out}_i\}_{i=0}^{d_{\rm out}^2-1}$ with $\sigma^{\rm in(out)}_i\in {\cal L}({\cal H}_{\rm in(out)})$, $\sigma^{\rm in(out)}_0:=\mathbb{1}_{\rm in(out)}$, and $\Tr(\sigma_i^{\rm in(out)} \sigma_j^{\rm in(out)})=d_{\rm in(out)}\delta_{i,j}$.
For any $\rho_{\rm in}\in{\cal S}({\cal H}_{\rm in})$, $\rho_{\rm out}\in{\cal S}({\cal H}_{\rm out})$, and $E \in {\cal E}({\cal H}_{\rm out})$ we introduce vectors $\boldsymbol{r}^{\rm in}(\rho_{\rm in}) \in \mathbb R^{d^2_{\rm in}-1}$, $\overline{\boldsymbol{r}}^{\rm in}(\rho_{\rm in}) \in \mathbb R^{d^2_{\rm in}}$, $\boldsymbol{r}^{\rm out}(\rho_{\rm out}) \in \mathbb R^{d^2_{\rm in}-1}$, and $\boldsymbol\eta^{\rm out}(E)\in \mathbb{R}^{d_{\rm out}^2-1}$ with corresponding elements
\begin{equation}
	\begin{aligned}
	&r^{\rm in}_i(\rho_{\rm in}) := \Tr (\sigma^{\rm in}_i \rho_{\rm in}), &i&=1,\ldots,d_{\rm in}^2-1;\\
        &\overline{r}^{\rm in}_i(\rho_{\rm in}) := \Tr (\sigma^{\rm in}_i \rho_{\rm in}^{\top}), &i&=0,\ldots,d_{\rm in}^2-1;\\
        &r^{\rm out}_i(\rho_{\rm out}):=\Tr (\sigma_i^{\rm out} \rho_{\rm out}), &i&=1,\ldots,d_{\rm out}^2-1;\\
        &\eta^{\rm out}_i(E):=\frac{1}{d_{\rm out}}\Tr (\sigma_i^{\rm out}E), &i&=1,\ldots,d_{\rm out}^2-1.
	\end{aligned}
\end{equation}
We note that $\overline{r}^{\rm in}_i(\rho_{\rm in})=d_{\rm in}^{-1}$ for every $\rho_{\rm in}$.
Then for a quantum channel $\Phi$ with Choi state $C_{\Phi}\in {\cal C}({\cal H}_{\rm in},{\cal H}_{\rm out})$ we can consider a real $(d_{\rm out}^2-1)\times {d_{\rm in}^2}$ matrix $\boldsymbol{C}(C_{\Phi})$ with elements
\begin{equation}
    C_{i,j}(C_{\Phi}):=\frac{1}{d_{\rm in}} \Tr \left(C_{\Phi} \sigma^{\rm in}_j \otimes \sigma^{\rm out}_i \right)
\end{equation}
(here $i=1,\ldots,d_{\rm out}^2-1$ and $j=0,\ldots,d_{\rm in}^2-1$).
It is clear that
\begin{eqnarray}
        &\boldsymbol{r}^{\rm out}(\Phi[\rho_{\rm in}]) &= \boldsymbol{C}(C_{\Phi}) \cdot \overline{\boldsymbol{r}}^{\rm in}(\rho_{\rm in})\quad \text{and} \label{eq:rout}\\
        &\Tr (E\Phi[\rho_{\rm in}]) &= \boldsymbol\eta^{\rm out}(E) \cdot \boldsymbol{r}^{\rm out}(\Phi[\rho_{\rm in}])+\eta_0^{\rm out}(E),\label{eq:etaout}
\end{eqnarray}
where we treat $\cdot$ in Eq.~\eqref{eq:rout} as matrix-vector multiplication, and $\eta_0^{\rm out}(E):=d_{\rm out}^{-1}\Tr (E)$.
Using Eq.~\eqref{eq:rout} we introduce a set
\begin{multline}
    \widetilde\Gamma_\text{\rm in-out}({\boldsymbol n}, \rho_{\rm in}, {\boldsymbol E}, \boldsymbol\epsilon) :=
    \left\{ \boldsymbol{C} \in \mathbb{R}^{d_{\rm out}^2-1} \times \mathbb{R}^{d_{\rm in}^2}:
    \right.\\
    \left.
    \boldsymbol{C} \cdot \overline{{\boldsymbol r}}^{\rm in}(\rho_{\rm in})\in \widetilde\Gamma_{\rm POVM}({\boldsymbol n}, {\boldsymbol E}, \boldsymbol\epsilon)
   \right\},
\end{multline}
which can be thought as an analog of $\Gamma_\text{\rm in-out}({\boldsymbol n}, \rho_{\rm in}, {\boldsymbol E}, \boldsymbol\epsilon)$ with removed  semipositivity condition (note, that the restriction on the partial trace remains).
Then we can also introduce a set
\begin{multline} \label{eq:proc_tilde}
   \widetilde\Gamma_{\rm CPTP}\left(\boldsymbol{\mathfrak{n}},
   \boldsymbol\rho_{\rm in},
   \boldsymbol{\mathfrak{E}},
   \boldsymbol{\varepsilon}\right)
   :=\\
    \bigcap_{i=1}^{M} \bigcap_{j=1}^{L^{(i)}}
   \widetilde\Gamma_\text{\rm in-out}\left({\boldsymbol n}^{(i)}_j,
   \rho^{(i)}_{\rm in},
   {\boldsymbol E}^{(i)}_j,
   \boldsymbol\epsilon^{(i)}_{j}\right),
\end{multline}
and see that by its construction
\begin{equation} \label{eq:incl-for-CPTP}
    C_\Phi \in \Gamma_{\rm CPTP}\left(
    \ldots
    \right)
   \Rightarrow \boldsymbol{C}(C_\Phi) \in \widetilde\Gamma_{\rm CPTP}\left(
   \ldots
   \right).
\end{equation}

One can see that $\widetilde\Gamma_{\rm CPTP}$ is a polyhedron in $\mathbb{R}^{d_{\rm in}^2(d_{\rm out}^2-1)}$, whose bounding planes are determined by input states and effects of corresponding output measurement POVMs.
As one may expect, the necessary and sufficient condition for $\widetilde\Gamma_{\rm CPTP}$ to be a polytope is information completeness of the QPT protocol.
\begin{thm}[Necessary and sufficient condition for QPT confidence polyhedron to be a polytope] \label{thm:CPTP-polytope}
	Consider a QPT protocol specified by a set of input states $\boldsymbol\rho_{\rm in}$,
	and corresponding measurements $\boldsymbol{\mathfrak{E}}$.
	For any appropriate measurement data $\boldsymbol{\mathfrak{n}}$ and appropriate-size tuple of positive real numbers $\boldsymbol\varepsilon$,
	$\widetilde\Gamma_{\rm CPTP}\left(\boldsymbol{\mathfrak{n}},
	\boldsymbol\rho_{\rm in},
	\boldsymbol{\mathfrak{E}},
	\boldsymbol{\varepsilon}\right)$ is a polytope (bounded polyhedron), iff
	$\boldsymbol\rho_{\rm in}$ and $\boldsymbol{\mathfrak{E}}$ yield informationally-complete QPT protocol, that is the matrix ${\boldsymbol A}(\boldsymbol\rho_{\rm in}, \boldsymbol{\mathfrak{E}})$, whose $(d_{\rm out}^2-1)d_{\rm in}^2$-dimensional rows consist of elements of $\overline{\boldsymbol{r}}^{\rm in}(\rho_{\rm in}^{(i)}) \otimes \boldsymbol\eta^{\rm out}(E^{(i)}_{j,k})$ for all possible $i, j, k$, has a trivial kernel.
\end{thm}
See Appendix~\ref{app:thm:CPTP-polytope} for the proof.

Thus, by realizing informationally-complete measurements over each of input states from a proper spanning set, one can obtain a bounded confidence region in the space Choi states (space of quantum channels), 
regardless of the restrictions posed by the boundedness of the whole space of Choi states.

\section{Deriving confidence intervals for affine functions} \label{sec:LF}

\begin{figure*}
	\begin{minipage}[h]{0.41\linewidth}
		\centering QST
		\includegraphics[width=\linewidth]{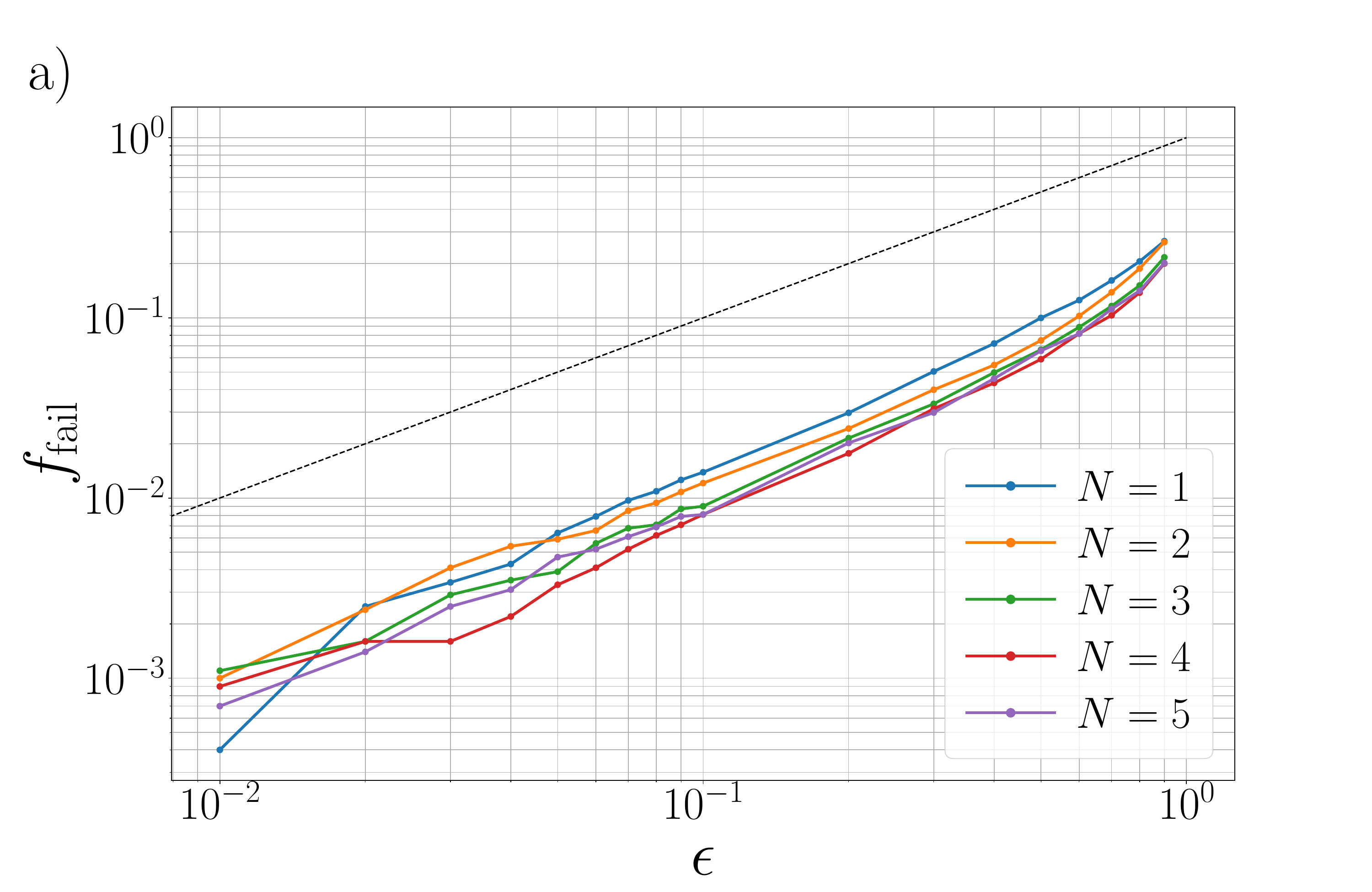}
		\includegraphics[width=\linewidth]{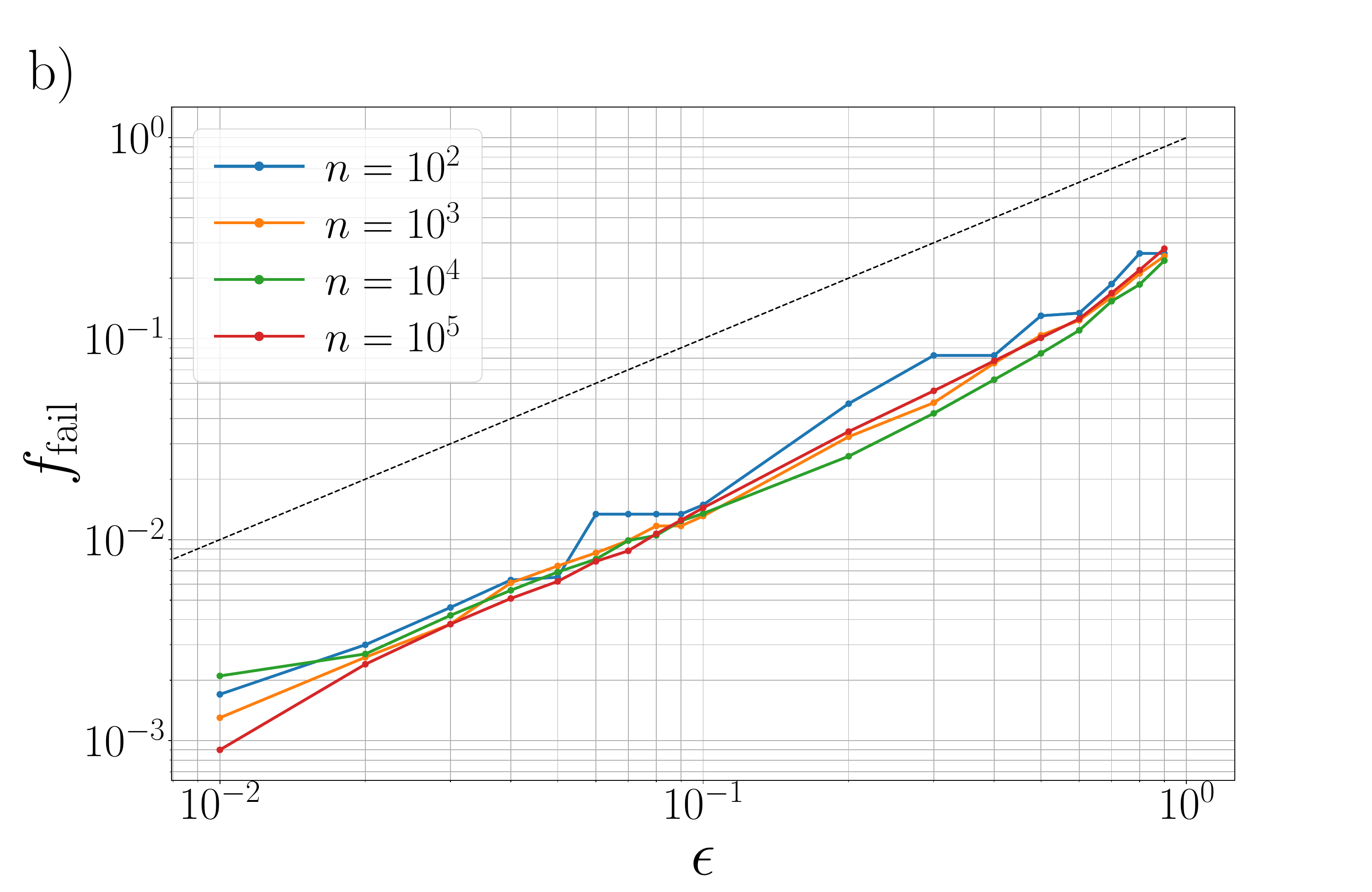}
	\end{minipage}readout
	\begin{minipage}[h]{0.41\linewidth}
		\centering QPT
 	       \includegraphics[width=\linewidth]{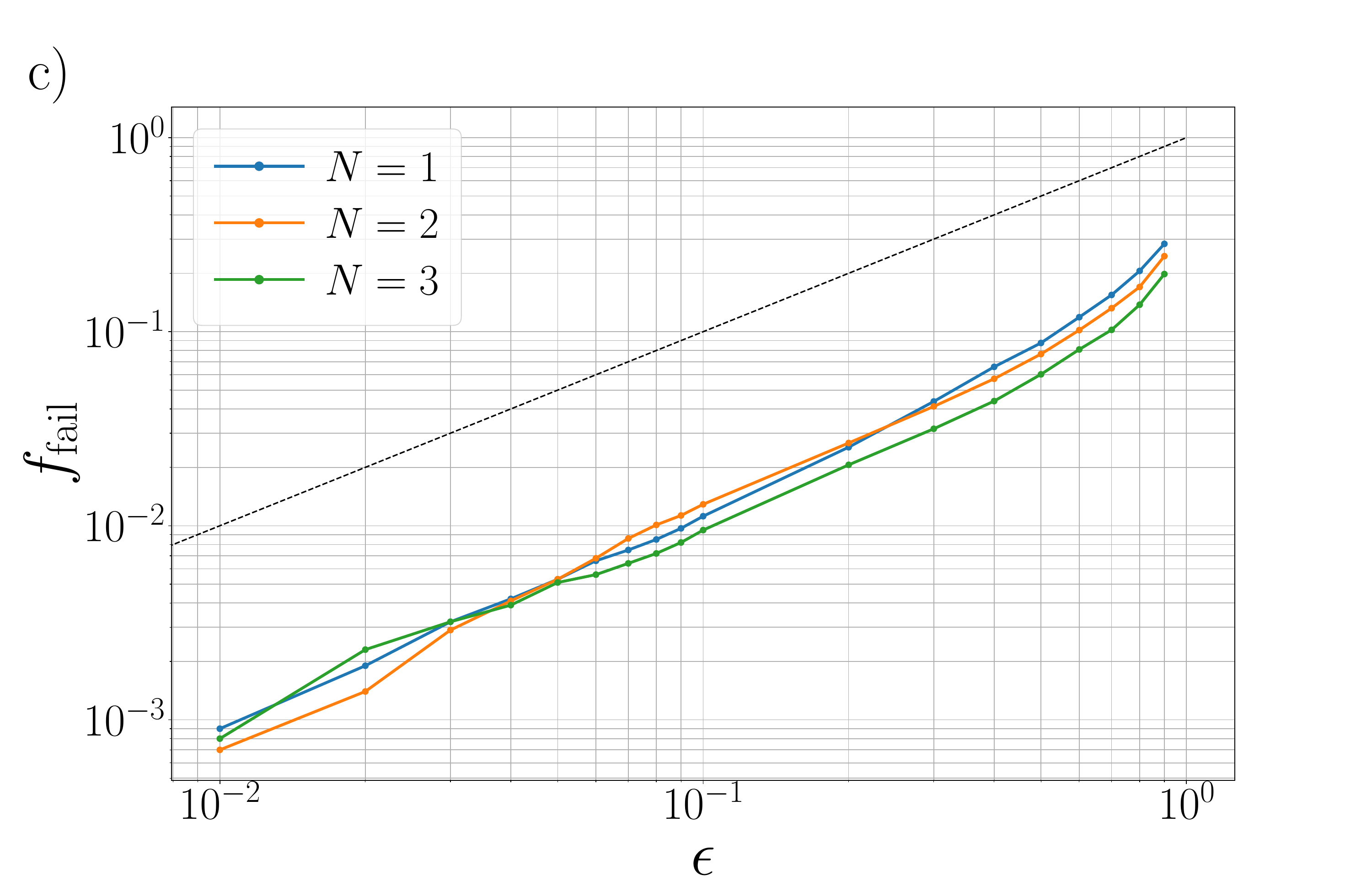}
		\includegraphics[width=\linewidth]{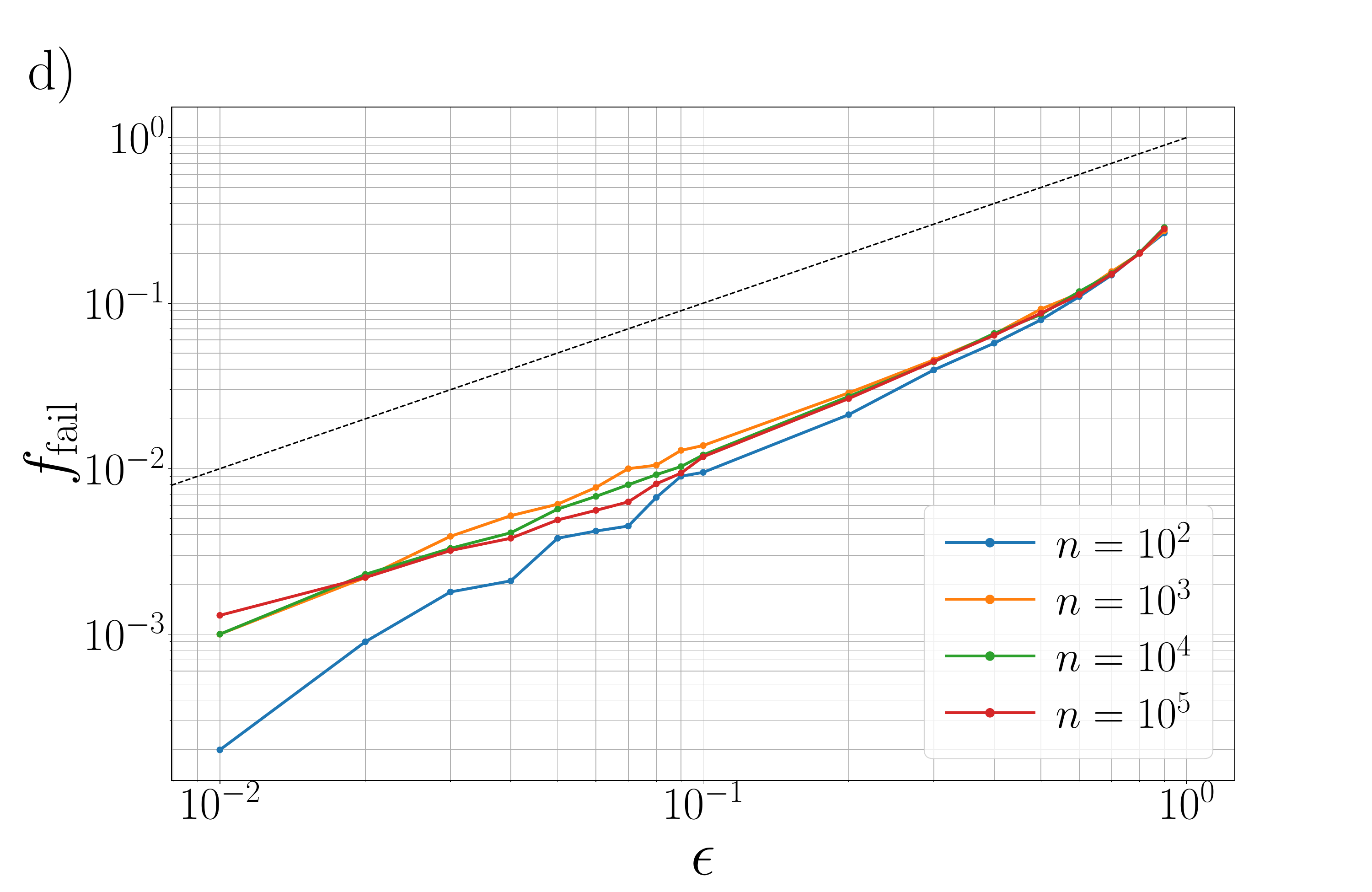}
	\end{minipage}
	\vskip-3mm
    \caption
    {
    Numerical results for the confidence polytopes performance:
    (a,b) results of the method for QST of $N$-qubit GHZ-type state $\ket{\psi^{(N)}}$ and (c,d) QPT of $N$-qubit depolarizing channel $\Phi^{(N)}_p$ with $p=0.1$.
    Out-of-confidence region events ratio $f_{\rm fail}$ is shown as a function of an upper bound on a probability of error $\epsilon$.
    Dashed line shows a critical level $f_{\rm fail}=\epsilon$.
    Each point is obtained from $10^4$ simulated tomography experiments, all $\epsilon_{i,j}$ and $\epsilon_{j,k}^{(i)}$ are considered to be equal to each other.
    In (a) and (c) the performance for different number of qubits $N$ and the fixed number of state/channel copies $n=10^5$ for each readout measurement (each pair of an input state and a readout measurement) is shown.
    In (b) and (d) the performance for different number state/channel copies $n$ for each readout measurement (each pair of an input state and a readout measurement) and the fixed number of qubits $N=1$ is depicted.
    }
    \label{fig:numerics}
\end{figure*}

\begin{figure*}
	\centering
	\includegraphics[width=0.45\linewidth]{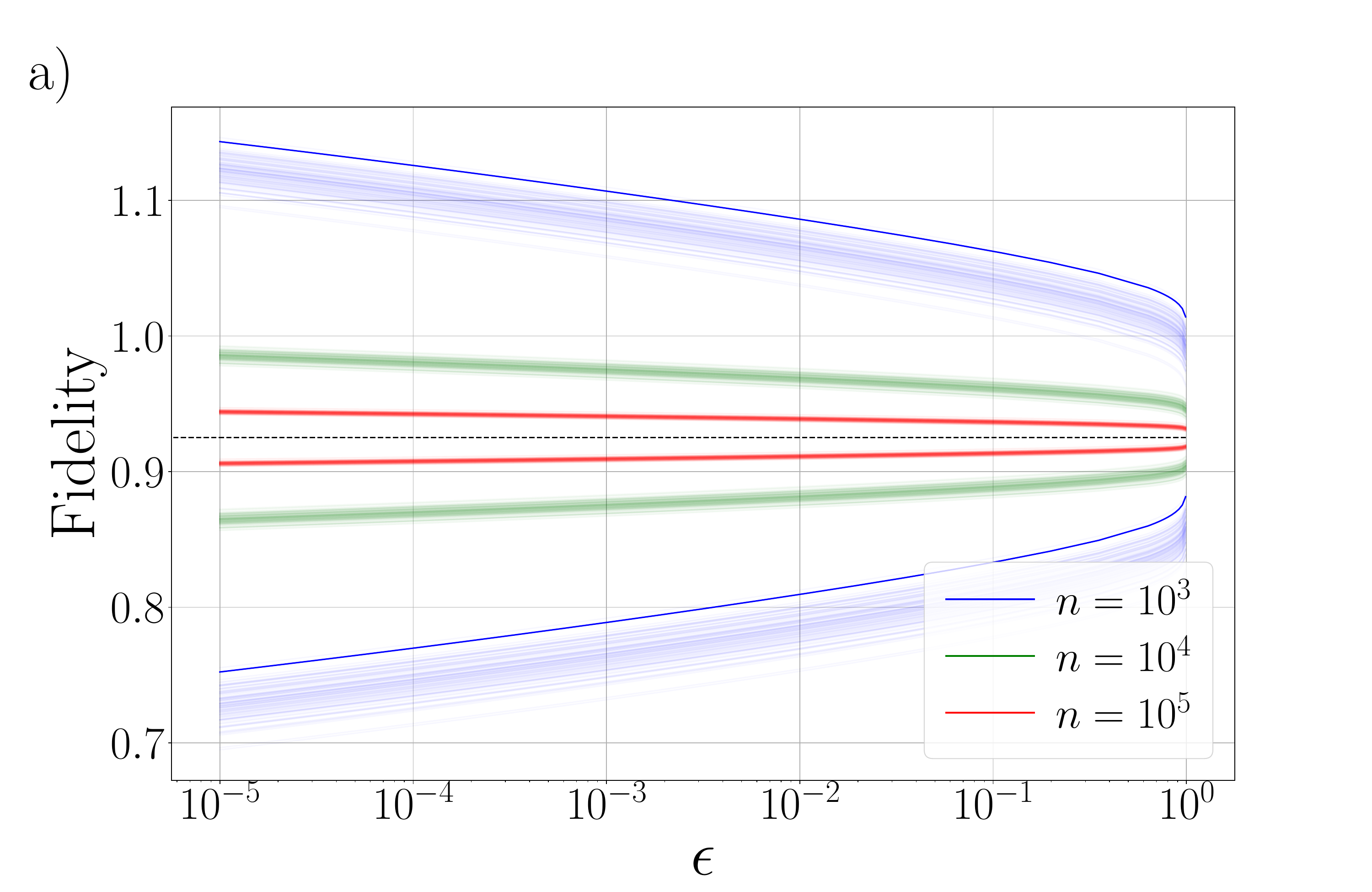}
	\includegraphics[width=0.45\linewidth]{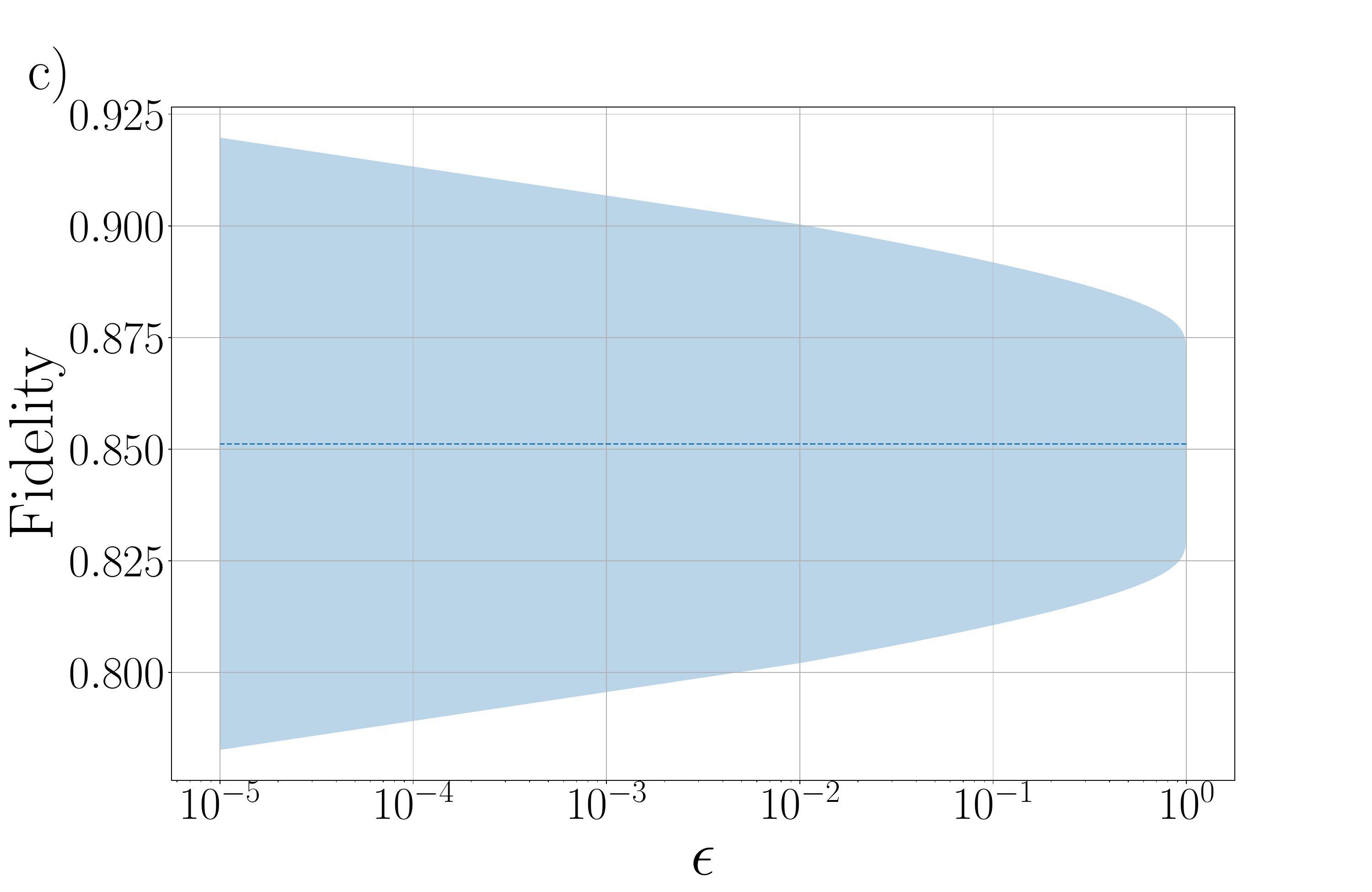}
	\includegraphics[width=0.8\linewidth]{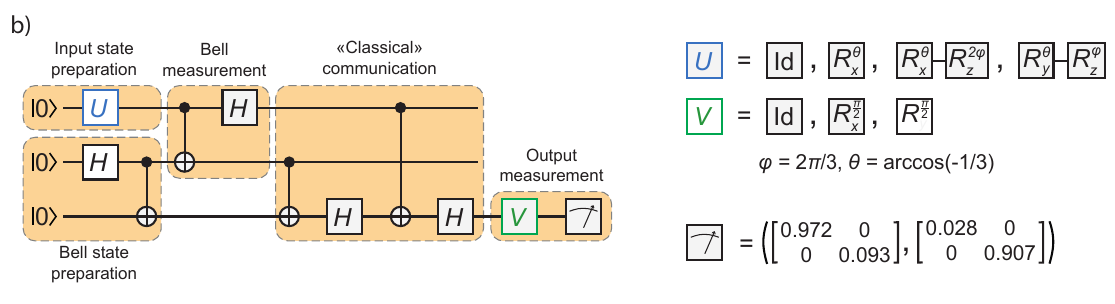}
	\vskip -5mm
	\caption{In (a) distribution of fidelity confidence intervals as the function of (one minus) confidence level over 100 numerical experiments for depolarizing channel and different values of $n$ is shown.
	The single confidence interval for $n=10^3$ is highlighted. The dashed line shows the true value of the fidelity.
	In (b) the circuit of the QPT protocol for the quantum teleportation channel is depicted (standard notations for Hadamard gate, Controlled-NOT gate, and computational basis readout measurement are used). 
	Gates $U$ and $V$ consistently take forms shown in the right-hand side, where $R_{x(y)}^{\bullet}$ is the standard rotation around $x(y)$ axis of the Bloch sphere (12 configurations of the circuit is run in total).
	The imperfections of readout computational basis measurement are also taken into account.
	In (c) the resulting fidelity confidence intervals for the QPT experiment from (b) run on IBM quantum experience superconducting processor is demonstrated.
	The dash line shows fidelity of the reconstructed point estimate of the channel.}
	\label{fig:fidelity}
\end{figure*}

In the previous sections, we considered the construction of confidence polytopes for unknown quantum states and quantum processes (channels).
Though these polytopes form rigorous confidence regions, they do not provide direct information about some values of practical interest, such as fidelity, mean values of observables, and so on.
The proposal of Ref.~\cite{Wang2019} is to use random sampling for generating a number of states inside confidence polytope and then extract confidence region for a value of interest by considering its minimum and maximum over generated states.
However, this approach may be quite computationally expensive, especially in the case of high-dimensional spaces, with which one deals in QPT.

Here we propose to use the linear structure of derived confidence regions and obtain confidence regions for linear functions by employing well-studied linear optimization tools~\cite{cvxopt}.
We consider linear optimization problem in the following form:
\begin{equation} \label{eq:LP}
    \begin{aligned}
        &\text{maximize~} {\boldsymbol k}\cdot {\boldsymbol x}\\
        &\text{s.t.~} {\boldsymbol A}{\boldsymbol x} \preceq {\boldsymbol b},
    \end{aligned}
\end{equation}
where ${\boldsymbol k}\in \mathbb{R}^\alpha$ is known vector for some $\alpha\geq1$,
${\boldsymbol A}=\begin{pmatrix} {\boldsymbol a}_1 & \ldots & {\boldsymbol a}_\beta\end{pmatrix}^\top$ is known $\alpha\times\beta$ real matrix for some $\beta\geq1$, ${\boldsymbol b}$ is known $\beta$-dimensional real vector,
${\boldsymbol x}\in \mathbb{R}^\alpha$ is unknown,
and $\boldsymbol{A}{\boldsymbol x} \preceq {\boldsymbol b}$ means that every element of vector $\boldsymbol{A}{\boldsymbol x}$ is not greater than corresponding element of ${\boldsymbol b}$.

Every `row' of the condition $\boldsymbol{A}{\boldsymbol x} \preceq {\boldsymbol b}$ can be written in the form ${\boldsymbol a}_i \cdot {\boldsymbol x} \leq b_i$, so we can represent $\boldsymbol{A}{\boldsymbol x} \preceq {\boldsymbol b}$ as
\begin{equation} \label{eq:generalpolytope}
    {\boldsymbol x} \in \Gamma, \quad \Gamma := \bigcap_{i=1}^\beta \Gamma_i, \quad
    \Gamma_i := \{{\boldsymbol x}\in\mathbb{R}^{\alpha}: {\boldsymbol a}_i \cdot {\boldsymbol x} \leq b_i\}.
\end{equation}
The structure of Eq.~\eqref{eq:generalpolytope} exactly coincides with the form of $\widetilde\Gamma_{\rm POVMs}$ and $\widetilde\Gamma_{\rm CPTP}$.
This fact opens the way to find confidence intervals for linear (or affine in general) functions $
\phi: {\cal S}({\cal H})\rightarrow \mathbb{R}$ and $\chi: {\cal C}({\cal H}_{\rm in}, {\cal H}_{\rm out})\rightarrow \mathbb R$ that can be represented as follows:
\begin{equation}
    \begin{aligned}
        \phi(\rho) &= {\boldsymbol r}(\rho)\cdot\boldsymbol\phi + \phi_0, \\
        \chi(C_\Phi) &=  {\boldsymbol c}(C_{\Phi})\cdot\boldsymbol\chi +\chi_0
    \end{aligned}
\end{equation}
correspondingly, 
where ${\boldsymbol c}(C_{\Phi}) \in \mathbb{R}^{d_{\rm in}^2(d_{\rm out}^2-1)}$ 
is a vector formed from stacked rows of ${\boldsymbol C}(C_{\Phi})$, $\boldsymbol\phi \in \mathbb{R}^{d^2-1}$,  $\boldsymbol\chi \in \mathbb{R}^{d_{\rm in}^2(d_{\rm out}^2-1)}$, and $\phi_0, \chi_0\in\mathbb{R}$

By obtaining solutions $\phi_{\max}'$ ($\phi_{\min}'$) of optimization problems
\begin{equation} \label{eq:QST-LP}
    \begin{aligned}
        &\text{maximize~} (-)\boldsymbol\phi \cdot {\boldsymbol r}\\
        &\text{s.t.~} {\boldsymbol r} \in \widetilde\Gamma_{\rm POVMs}(\vec{\boldsymbol n}, \vec{\boldsymbol E}, \vec{\boldsymbol\epsilon}),
    \end{aligned}
\end{equation}
or solutions $\chi_{\max}'$ ($\chi_{\min}'$) of
\begin{equation} \label{eq:CPTP-LP}
    \begin{aligned}
        &\text{maximize~} (-) \boldsymbol\chi \cdot {\boldsymbol c}\\
        &\text{s.t.~} {\boldsymbol c} \in \widetilde\Gamma_{\rm CPTP}\left(\boldsymbol{\mathfrak{n}},
   \boldsymbol\rho_{\rm in},
   \boldsymbol{\mathfrak{E}},
   \boldsymbol\varepsilon\right)
    \end{aligned}
\end{equation}
one can reconstruct confidence intervals for the values of $\phi(\rho)$ and $\chi(C)$ in the corresponding forms:
\begin{equation}
\begin{aligned}
    \Pr{\phi(\rho)-\phi_0\in[\phi'_{\min},\phi'_{\max}]} &> 1-{\rm CL}_{\rm POVMs}(\vec{\boldsymbol\epsilon}),\\
    \Pr{\chi(C)-\chi_0\in[\chi'_{\min},\chi'_{\max}]} &> 1-{\rm CL}_{\rm CPTP}(\boldsymbol\varepsilon),
\end{aligned}
\end{equation}
We note that the existence of solution of Eq.~\eqref{eq:QST-LP} and Eq.~\eqref{eq:CPTP-LP} is guaranteed only in the case of boundness of the corresponding polyhedron.

The variety of possible functions $\phi$ includes but is not limited to fidelity with respect to pure state $\ket{\psi}$ ($\phi(\rho)=\bra{\psi}\rho\ket{\psi}$), 
mean value of an observable $O$ ($\phi(\rho)=\Tr (\rho O)$), and probability of an outcome corresponding for some effect $E$ ($\phi(\rho)=\Tr (\rho E)$).
List of possible functions $\chi$ includes fidelity with respect to unitary channel $C_{U}$ ($\chi(C)=d_{\rm in}^{-2}\Tr(CC_U)$), 
mean value of an observable $O_{\rm out}$ for output state given input state $\rho_{\rm in}$ ($\chi(C)=\Tr (\rho_{\rm in}^{\top} \otimes O_{\rm out} C))$, 
and probability of effect $E_{\rm out}$ for output state given input state $\rho_{\rm in}$ ($f(C)=\Tr (\rho_{\rm in}^{\top} \otimes E_{\rm out} C)$).
We note that these functions have a clear physical meaning, and the proposed method allows one to obtain guaranteed-precision estimates on the values for these functions.
Thus, our method is then can be used in experiments on benchmarking quantum information processing devices.
We provide some particular examples in the next section.

\section{Performance analysis} \label{sec:performance}

An important question is the practical applicability of the proposed method in practical settings.
Here we analyze the performance of the confidence polytopes one the basis of numerical simulation and experimental data from IBM cloud superconducting processor. 
To provide the full picture, we consider both QST and QPT cases. 
We note that the results for QST can be also found in the original paper on polytopes for QST~\cite{Wang2019}.

For the QST setting, we consider a Greenberger-Horne-Zeilinger (GHZ) state of $N$ qubits of the following form:
\begin{equation}
    \ket{\psi^{(N)}}=\frac{1}{\sqrt{2}}( \ket{0}^{\otimes N}+\ket{1}^{\otimes N}),
\end{equation}
and base our tomographic reconstruction on $3^N$ POVMs corresponding to single-qubit protective measurements with respect to $x$, $y$, and $z$ axes.
We consider the total number $n$ copies of $\ket{\psi^{(N)}}$
for each of  $3^N$ variants of readout measurements.

In the QPT case, we consider an $N$-qubit depolarizing channel
\begin{equation}
	\Phi^{(N)}_p[\rho] = (1-p)\rho + \frac{\mathbb{1}}{2^N}\Tr \rho,
\end{equation}
where the $\mathbb{1}$ is $2^N \times 2^N$ identity matrix.
In our experiments we fix $p=0.1$.
The set of $4^N$ input states consists of pure product states in the following form:
\begin{equation}
    \rho_{\rm in}^{(i_1,\ldots,i_N)} = \rho_{\rm in}^{(i_1)}\otimes\ldots\otimes\rho_{\rm in}^{(i_N)},
\end{equation}
where $i_j\in\{1,\ldots,4\}$ and single-qubit states $\{\rho_{\rm in}^{(k)}\}_{k=1}^4$ form a tetrahedron inscribed in a cube with facet orthogonal to $x$, $y$, and $z$ axes.
The output of each of $4^N$ input states is then measured with $3^N$ POVMs in the same way as in the QST case.
For each configuration of input state and readout measurement $n$ copies of $\Phi^{(N)}_p[\cdot]$ are considered.

For each QST and QPT simulations, we reconstruct confidence regions with respect to different confidence levels and check whether the true state or channel falls into this region.
While constructing confidence regions and confidence levels, we set all $\epsilon_{i,j}$ and $\epsilon^{(i)}_{j,k}$ to be equal to each other.
After repeating the experiment several times, we count fractions of experiments $f_{\rm fail}$ for which the true state or channel appeared to be outside the confidence region.
Of course, by definition of a confidence level, we expect that $f_{\rm fail}<\epsilon$, where $\epsilon:=1-{\rm CL}_{\rm POVMs}(\vec{\boldsymbol\epsilon})$ (for QST) and $\epsilon:=1-{\rm CL}_{\rm CPTP}(\boldsymbol\varepsilon)$ (for QPT).
On the other hand, the distance between $f_{\rm fail}$ and $\epsilon$ characterizes the tightness of the reconstructed confidence region.

The obtained numerical results are shown in Fig.~\ref{fig:numerics}.
First of all, we note that both in QST and QPT cases, $f_{\rm fail}$ as a function of $\epsilon$ has similar behavior for different values of $n$ and $N$.
The only a bit out of the ordinary behavior is demonstrated for QST confidence polytopes with $n=10^{-2}$ and $\epsilon \lesssim 10^{-1}$, though the corresponding behavior for larger considered values of $n$ is almost the same.
In this way, the general conclusion about the scalability of confidence polytopes reported for QST in Ref.~\cite{Wang2019} can be generalized to the case of quantum channels.
Second, we see that $f_{\rm fail}$ is usually approximately one order of magnitude less than $\epsilon$.
This overestimation seems to be a price for the rigorousness of the employed Clopper-Pearson confidence intervals. 
We note that the safeness of the developed approach can be utilized, e.g. in the framework of quantum cryptography, where the fairness of the used bounds is of critical importance in the certification of such devices.

We then consider the construction of the confidence region for the affine function of the Choi state, in particular the fidelity of the quantum channel with respect to a unitary process.
As an example, we consider a single-qubit depolarizing channel $\Phi[\cdot]:=\Phi_{0.1}^{(1)}[\cdot]$, which is treated as unknown process and estimate fidelity with respect to ideal identical channel as follows:
\begin{equation}
	\phi(C) = \frac{1}{2}\bra{\psi^{(2)}} C \ket{\psi^{(2)}},
\end{equation}
where $C$ is the Choi matrix of the certain qubit-to-qubit channel (here we use the fact that Choi state for the identical qubit-to-qubit channel is given by $2\ket{\psi^{(2)}}\bra{\psi^{(2)}})$.

The results of the reconstruction of confidence interval $(\phi_{\min}, \phi_{\max})$ for $\phi(C)$ 
based on simulated experiment data generated for different values of $n$ with respect to $\Phi[\cdot]$ are shown in Fig.~\ref{fig:fidelity}a (as in the QPT experiments described before, we set all $\epsilon^{(i)}_{j,k}$ to be equal to each other).
We see that the obtained confidence regions are rigorous and very accurate: even for $\epsilon=0.5$ in all simulated experiments the true fidelity $\phi(C_{\Phi})=1-3p/4$ is well inside the corresponding confidence region.
One can also see that for $n=10^3$ the upper bound $\phi_{\max}$ can be larger than one, that is the consequence of the fact that $\widetilde\Gamma_{\rm CPTP}$ is larger than the space of physical plausible quantum channels.

We also consider a realistic QPT experiment performed at the superconducting quantum processor provided by IBM.
The circuit of the experiment corresponds to the quantum teleportation of a state of the 1st qubit on the 3rd qubit (see Fig.~\ref{fig:fidelity}b).
In the ideal case, this process coincides with the identical channel.
In Fig.~\ref{fig:fidelity}c we show the behavior of the confidence region of the fidelity with respect to the identical qubit-to-qubit channel for the case of 
$n=2^{13}\approx 8\times10^3$ experiments for each configuration of an input state and a readout measurement.
We see that the results of the methods can be used as lower and upper bounds on the performance characteristics of existing quantum devices with the possibility to use this approach for the noisy intermediate-scale quantum (NISQ) machines.

\section{Conclusion}\label{sec:conclusion}

In the present paper we have provided the generalization of the QST confidence polytopes approach, based on Clopper-Pearson intervals, to the QPT scenario.
Specifically, we have derived confidence regions in the form of a polyhedron (polytope) for a Choi matrix of an unknown quantum process (channel) based on the measurement results of output states given known input ones. 
Then we have shown how QST (QPT) confidence polytopes can be used for extracting confidence intervals for affine functions of quantum states (Choi states) with standard methods of linear programming.

Our numerical experiments have indicated on the scalability of the our approach with increasing the dimensionality of the process and the amount of processed measurement data.
Moreover, the results obtained with the IBM cloud quantum processor have shown the applicability of our approach for characterizing NISQ devises.

The comparison of confidence levels with the actual frequency of falling true density matrices (Choi states) into confidence polytopes have shown that the obtained values of confidence levels are quite conservative: 
the estimated real fail probability is commonly one order less than the one obtained from the confidence level.
This fact can be considered as a price of rigorousness of all the derivations, including the derivation of Clopper-Pearson intervals.
Nevertheless, the safeness of the derived confidence regions and confidence intervals can be extremely useful in situations, where the accuracy of the reconstructed upper and lower bounds is of particular importance, 
such as quantum key distribution.

On the practical side, we also note that that a straightforward approach to QPT, i.e. probing the process with a set of states, whose density operators form a spanning set in the space of all operators over a particular Hilbert space, 
may be challenging~\cite{Poyatos1997}. 
One of the possible solutions is to use coherent states as probes for QPT~\cite{Lobino2008,Keshari2011,Anis2012,Fedorov2015}.
A direction for the further research is then to adopt our approach for the case coherent-state QPT protocols.

\section*{Acknowledgements}
We are grateful to A. Lvovsky, M. Ringbauer, and E. Tiunov for fruitful discussion. 
We acknowledge use of the IBM Q Experience for this work. 
The views expressed are those of the authors and do not reflect the official policy or position of IBM or the IBM Q Experience team.
This work is supported by the Russian Science Foundation (Grant No. 20-42-05002; modification of the confidence polytopes approach in Sec~\ref{sec:QST} and analysis of the application to real quantum computing devices in Sec.~\ref{sec:performance}).
The authors also acknowledge the support of Leading Research Center on Quantum Computing (Agreement No. 014/20; development of the method in Secs.~\ref{sec:QPT}-\ref{sec:LF}).

\appendix

\section{Theorems proofs}

\subsection{Proof of Theorem~\ref{thm:POVMs}}  \label{app:thm:POVMs}
\begin{proof}
	In order to prove the Theorem, we show that 
	\begin{equation} \label{eq:CR_for_POVMs}
		\sum_{\vec{\boldsymbol n}}\Pr{\vec{\boldsymbol n}|\rho,\vec{\boldsymbol E}} \gamma(\rho\in\Gamma_{\rm POVMs}) > {\rm CL}(\vec{\boldsymbol\epsilon}),
	\end{equation}
	where $\Pr{\vec{\boldsymbol n}|\rho,\vec{\boldsymbol E}}$ is the probability to obtain measurement data $\vec{\boldsymbol n}=({\boldsymbol n}_i)_{i=1}^L$ after measuring $\rho$ 
	with the set of POVMs $\vec{\boldsymbol E}=({\boldsymbol E}_i)_{i=1}^L$ and $\gamma(\cdot)$ is an indicator function. 
	Further we note that $\Pr{\vec{\boldsymbol n}|\rho,\vec{\boldsymbol E}}$ can be factorized due to the fact measurement outcomes related to different POVMs are independent:
	\begin{equation}
		\Pr{\vec{\boldsymbol n}|\rho,\vec{\boldsymbol E}}=\prod_{i=1}^L \Pr{{\boldsymbol n}_i|\rho,{\boldsymbol E}_i}.
	\end{equation}
	We can also factorize the indicator function:
	\begin{multline}
		\gamma\left(\rho\in\Gamma_{\rm POVMs}(\vec{\boldsymbol n}, \vec{\boldsymbol E}, \vec{\boldsymbol\epsilon})\right)=\\\prod_{i=1}^L 
		\gamma\left(\rho\in\Gamma_{\rm POVM}({\boldsymbol n}_i,{\boldsymbol E}_i,\boldsymbol\epsilon_i)\right).
	\end{multline}
	As the result, we obtain
	\begin{multline}
		\sum_{{\vec{\boldsymbol n}}}\Pr{\vec{\boldsymbol n}|\rho,\vec{\boldsymbol E}} \gamma\left(\rho\in\Gamma_{\rm POVMs}\right) \\
		=\prod_{i=1}^{L}\sum_{{\boldsymbol n}_i}\Pr{{\boldsymbol n}_i|\rho, {\boldsymbol E}_i}\gamma\left(\rho\in\Gamma_{\rm POVM}({\boldsymbol n}_i,{\boldsymbol E}_i,\boldsymbol\epsilon_i)\right).
    \end{multline}
    According to Theorem~\ref{thm:Renner},
	\begin{multline}
		\sum_{{\boldsymbol n}_i}\Pr{{\boldsymbol n}_i|\rho, {\boldsymbol E}_i}\gamma\left(\rho\in\Gamma_{\rm POVM}({\boldsymbol n}_i,{\boldsymbol E}_i,\boldsymbol\epsilon_i)\right) \\> 
		{\rm CL}_{\rm POVM}(\boldsymbol\epsilon_i)=1-\sum_{j=1}^{P_i}\epsilon_{i,j},
	\end{multline}
	and so we have approached the desired relation~\eqref{eq:CR_for_POVMs}.
\end{proof}

\subsection{Proof of Theorem~\ref{thm:POVMs-polytope}} \label{sec:thm:POVMs-polytope}
\begin{proof}
	First of all, we note that the boundness of a polyhedron $\widetilde\Gamma_{\rm POVMs}(\vec{\boldsymbol n},\vec{\boldsymbol E},\vec{\boldsymbol\epsilon})$ 
	is equivalent to the fact that for any nonzero $\boldsymbol{v}\in \mathbb{R}^{d^2-1}$ there exists an effect $E_{i,j}$ within the set of POVMs $\vec{\boldsymbol E}$ such that $\boldsymbol{v}\cdot \boldsymbol\eta(E_{i,j})>0$.
	This is because if one were to move from a point $\boldsymbol{r}\in \widetilde\Gamma_{\rm POVMs}(\vec{\boldsymbol n},\vec{\boldsymbol E},\vec{\boldsymbol\epsilon})$ 
	in the direction $\boldsymbol{v}$, then starting from some nonnegative $\lambda'$ it would appear that
	\begin{equation}
		(\boldsymbol{r}+\lambda \boldsymbol{v}) \cdot \boldsymbol\eta(E_{i,j}) > \frac{n}{N}+\delta_N(n,\epsilon_{i,j})-\eta_0(E_{i,j}), 
	\end{equation}
	for $\lambda>\lambda'$. 
	That is the point $\boldsymbol{r}+\lambda\boldsymbol{v}$ is outside $\widetilde\Gamma_{\rm POVMs}(\vec{\boldsymbol n},\vec{\boldsymbol E},\vec{\boldsymbol\epsilon})$.
    
	Let us then prove that if $\vec{\boldsymbol E}$ is informationally complete, then $\widetilde\Gamma_{\rm POVMs}(\vec{\boldsymbol n},\vec{\boldsymbol E},\vec{\boldsymbol\epsilon})$ is bounded.
	The proof is by contradiction.
	Let there exist nonzero $\boldsymbol{v}\in \mathbb{R}^{d^2-1}$, such that 
	\begin{equation}
		\boldsymbol{v}\cdot \boldsymbol\eta(E_{i,j})\leq 0, 
		\quad i=1,\ldots,L,
		\quad j=1,\ldots,P_i.
	\end{equation}
	Then, for each $i$ we have
	\begin{equation}
	\begin{split}
		0\geq \boldsymbol{v}\cdot\boldsymbol\eta(E_{i,P_i}) &= \boldsymbol{v}\cdot \left( \boldsymbol\eta(\mathbb{1})-
 	       \sum_{j=1}^{P_i-1} \boldsymbol\eta(E_{i,j}) \right)\\
	        &=
	        -\sum_{j=1}^{P_i-1} \boldsymbol{v} \cdot \boldsymbol\eta(E_{i,j})\geq 0,
	\end{split}
	\end{equation}
	where we used the standard normalization condition for POVM effects.
   	 So we conclude that $\boldsymbol{v}\cdot \boldsymbol\eta(E_{i,j})=0$ for every $i$ and $j$. 
   	 However, this observation contradicts information completeness of $\vec{\boldsymbol E}$: 
	 One can find two distinct states $\rho^{(\alpha)}=\frac{1}{d}( \mathbb{1}+\lambda^{(\alpha)}\sum_{k=0}^{d^2-1}\sigma_k v_k)$, with $\alpha=1,2$ 
	 and scaling factors $\lambda^{(1)}\neq \lambda^{(2)}$ chosen to assert $\rho^{(\alpha)}\in{\cal S}({\cal H})$, that provide exactly the same measurement statistics.
    
	To prove that for an informationally \emph{incomplete} set $\vec{\boldsymbol E}$, the corresponding polyhedron $\widetilde\Gamma_{\rm POVMs}(\vec{\boldsymbol n},\vec{\boldsymbol E},\vec{\boldsymbol\epsilon})$ is unbounded, 
	we just note that there have to exist two distinct states $\rho^{(\alpha)}\in{\cal S}({\cal H})$ ($\alpha=1,2$) which provide the same statistics. 
	According to Eq.~\eqref{eq:probofmeasinvec}, for $\boldsymbol{v}:=\boldsymbol{r}(\rho^{(1)}-\rho^{(2})$ one has
	$\boldsymbol{v} \cdot \boldsymbol\eta(E_{i,j})=0$ for every $i$ and $j$, and hence $\widetilde\Gamma_{\rm POVMs}(\vec{\boldsymbol n},\vec{\boldsymbol E},\vec{\boldsymbol\epsilon})$ is unbounded.
\end{proof}

\subsection{Proof of Theorem~\ref{thm:CPTP}} \label{app:thm:CPTP}
\begin{proof}
	The proof is based on the observation that the probability $\Pr{{\boldsymbol{\mathfrak{n}}}|\Phi,\boldsymbol\rho_{\rm in},{\boldsymbol{\mathfrak{E}}}}$ 
	of obtaining measurement data $\boldsymbol{\mathfrak{n}}$ for an unknown CPTP map $\Phi$ by employing input states $\boldsymbol\rho_{\rm in}$ and output measurements $\boldsymbol{\mathfrak{E}}$, can be factorized as follows:
	\begin{equation}
		\Pr{\boldsymbol{\mathfrak{ n}}|\Phi,\boldsymbol\rho_{\rm in},\boldsymbol{\mathfrak{ E}}}=\prod_{i=1}^M\prod_{j=1}^{L^{(i)}}\Pr{{\boldsymbol n}^{(i)}_j|\rho_{\rm out}^{(i)},E^{(i)}_j},
	\end{equation}
	where $\rho_{\rm out}^{(i)}:=\Phi[\rho_{\rm in}^{(i)}]$.
	At the same time, the correponding indicator function can be factorized as
	\begin{equation}
		\gamma\left(C_\Phi\in\Gamma_{\rm CPTP}(\boldsymbol{\mathfrak{ n}},\vec{\boldsymbol\rho}_{\rm in},\boldsymbol{\mathfrak{ E}},\boldsymbol\varepsilon)\right)=\prod_{i=1}^M\prod_{j=1}^{L^{(i)}}\gamma(C_\Phi\in \Gamma^{(i)}_j),
	\end{equation}
	where
	\begin{multline}
		\Gamma^{(i)}_j := \left\{C\in{\cal C}({\cal H}_{\rm in}, {\cal H}_{\rm out}): \phantom{n^{(i)}_j} \right. \\
		\left. \Tr_{\rm in}(\rho_{\rm in}^{(i)\top}\otimes \mathbb{1}_{\rm out} C) \in \Gamma_{\rm POVM}\left({\boldsymbol n}^{(i)}_j,E^{(i)}_j, \boldsymbol\epsilon^{(i)}_j\right) \right\}.
	\end{multline}
	Since
	\begin{equation}
		C_\Phi \in  \Gamma^{(i)}_j 
		\Leftrightarrow
		\rho^{(i)}_{\rm out}\in \Gamma_{\rm POVM}({\boldsymbol n}^{(i)}_j,E^{(i)}_j, \boldsymbol\epsilon^{(i)}_j),
	\end{equation}
	we can write
	\begin{multline}\label{eq:CL-CPTP-pred}
		\sum_{\boldsymbol{\mathfrak{n}}}\Pr{\boldsymbol{\mathfrak{n}}|\Phi,\boldsymbol\rho_{\rm in},\boldsymbol{\mathfrak{E}}}\gamma(C_\Phi\in\Gamma_{\rm CPTP}
		(\boldsymbol{\mathfrak{n}},\boldsymbol\rho_{\rm in},\boldsymbol{\mathfrak{E}}, \boldsymbol\varepsilon)) 
		\\
		=\prod_{i=1}^M\prod_{j=1}^{L^{(i)}}\sum_{{\boldsymbol n}^{(i)}_{j}}
		\Pr{{\boldsymbol n}^{(i)}_j|\rho_{\rm out}^{(i)},\boldsymbol{E}^{(i)}_j}\\
		\times\gamma\left(\rho_{\rm out}^{(i)}(C)\in \Gamma_{\rm POVM}\left({\boldsymbol n}^{(i)}_j,\boldsymbol{E}^{(i)}_j,\boldsymbol\epsilon^{(i)}_{j}\right)\right).
	\end{multline}
	According to Theorem~\ref{thm:CPTP}, we obtain
	\begin{multline}\label{eq:POVMinCPTP}
		\sum_{{\boldsymbol n}^{(i)}_{j}}
		\Pr{{\boldsymbol n}^{(i)}_j|\rho_{\rm out}^{(i)},\boldsymbol{E}^{(i)}_j}\\
		\times\gamma\left(\rho_{\rm out}^{(i)}(C)\in \Gamma_{\rm POVM}\left({\boldsymbol n}^{(i)}_j,\boldsymbol{E}^{(i)}_j,\boldsymbol\epsilon^{(i)}_{j}\right)\right)
		\\ > {\rm CL}_{\rm POVM}\left(\boldsymbol\epsilon^{(i)}_j\right) = 1-\sum_{k=1}^{P^{(i)}_{j,k}}\epsilon^{(i)}_{j,k}.
	\end{multline}
	Substituting Eq.~\eqref{eq:POVMinCPTP} in Eq.~\eqref{eq:CL-CPTP-pred} gives us
	\begin{multline}
		\sum_{{\boldsymbol n}^{(i)}_{j}}
		\Pr{{\boldsymbol n}^{(i)}_j|\rho_{\rm out}^{(i)},\boldsymbol{E}^{(i)}_j}\\
		\times\gamma\left(\rho_{\rm out}^{(i)}(C)\in \Gamma_{\rm POVM}\left({\boldsymbol n}^{(i)}_j,\boldsymbol{E}^{(i)}_j,\boldsymbol\epsilon^{(i)}_{j}\right)\right)
		\\
		>\prod_{i=1}^{M}\prod_{j=1}^{L^{(i)}}\left(1-\sum_{k=1}^{P^{(i)}_{j}}\epsilon^{(i)}_{j,k}\right)={\rm CL}_{\rm CPTP}\left(\boldsymbol\varepsilon\right)
	\end{multline}
	that is the definition of the confidence level.
\end{proof}

\subsection{Proof of Theorem~\ref{thm:CPTP-polytope}} \label{app:thm:CPTP-polytope}
\begin{proof}
	We first note that in the view of Eqs.~\eqref{eq:rout} and~\eqref{eq:etaout}, each facet of the polyhedron $\widetilde\Gamma_{\rm CPTP}$ is given by the following equation:
	\begin{equation}
		\left({\boldsymbol r}^{\rm in}(\rho_{\rm in}^{(i)})\otimes \boldsymbol\eta(E^{(i)}_{j,k})\right) \cdot {\boldsymbol c} \leq \frac{n^{(i)}_{j,k}}{n^{(i)}_{j}} + \delta_{n^{(i)}_j}
	\end{equation}
	with $n^{(i)}_j:=\sum_k n^{(i)}_{j,k}$.
    
	To prove that the boundness $\widetilde\Gamma_{\rm CPTP}$ forces ${\rm ker}{\boldsymbol A}(\boldsymbol\rho_{\rm in}, \boldsymbol{\mathfrak{E}})$ to be trivial, 
	we note that otherwise, one can choose a nonzero ${\boldsymbol c}'\in {\rm ker}{\boldsymbol A}(\boldsymbol\rho_{\rm in}, \boldsymbol{\mathfrak{E}})$ and for each $\lambda\in\mathbb{R}$, 
	${\boldsymbol c}\in \widetilde\Gamma_{\rm CPTP}(\boldsymbol{\mathfrak{n}},\boldsymbol{\rho}_{\rm in},\boldsymbol{\mathfrak{E}},\vec{\boldsymbol\epsilon})$,
	\begin{equation}
		{\boldsymbol c}+\lambda{\boldsymbol c}' \in \widetilde\Gamma_{\rm CPTP}(\boldsymbol{\mathfrak{n}},\boldsymbol{\rho}_{\rm in},\boldsymbol{\mathfrak{E}},\vec{\boldsymbol\epsilon}),
	\end{equation}
	that contradicts the boundness assumption.
    
	The proof that the triviality of ${\rm ker}{\boldsymbol A}(\boldsymbol\rho_{\rm in}, \boldsymbol{\mathfrak{E}})$ yields the boundness of $\widetilde\Gamma_{\rm CPTP}$ is also by contradiction.
	Suppose that for some nonzero ${\boldsymbol c}' \in \mathbb{R}^{(d_{\rm out}^2-1)d_{\rm in}^2}$ 
	\begin{equation} \label{eq:leq0assumption}
		\left({\boldsymbol r}^{\rm in}(\rho_{\rm in}^{(i)})\otimes \boldsymbol\eta(E^{(i)}_{j,k})\right) \cdot {\boldsymbol c}' < 0
	\end{equation}
	for each $i,j,k$.
	However, from the fact that $\sum_k E^{(i)}_{j,k}=\mathbb{1}_{\rm out}$, it follows that $\sum_k \boldsymbol\eta^{\rm out}(E^{(i)}_{j,k})=0$.
	Therefore,
	\begin{equation}
		\sum_{i,j,k}\left({\boldsymbol r}^{\rm in}\left(\rho_{\rm in}^{(i)}\right)\otimes \boldsymbol\eta(E^{(i)}_{j,k})\right) \cdot {\boldsymbol c} = 0
	\end{equation}
	that contradicts with the initial assumption~\eqref{eq:leq0assumption}.
\end{proof}

{\bibliographystyle{unsrt}}

\end{document}